\documentclass[onecolumn]{IEEEtran}
\IEEEoverridecommandlockouts

\usepackage{graphics,graphicx} 

\usepackage{amsmath}
\usepackage{amssymb}
\usepackage{tikz}
\usepackage{amsthm}
\usepackage{comment}
\usepackage{mathcomSTEP}
\usepackage{subcaption}
\usepackage[utf8]{inputenc}
\usepackage{bm}

\usepackage{tikz}
\usetikzlibrary{positioning}
\usetikzlibrary{arrows,fit}

\usepackage{epstopdf}
\usepackage{pgfplots,setspace}

\newtheorem{prop}[thm]{Proposition}
\newtheorem{lemma}[thm]{Lemma}
\newtheorem{rem}[thm]{Remark}

\newcommand{\mmse}{\mathsf {mmse}}
\newcommand{\tr}{\mathrm {Tr}}
\newcommand{\Enc}{\mathrm {Enc}}
\newcommand{\rank}{\mathrm {rank}}
\newcommand{\Roo}{\overline{R}}

\tikzstyle{int}=[draw, fill=blue!10, minimum height = 1cm, minimum width=1.5cm,thick ]
\tikzstyle{sint}=[draw, fill=blue!10, minimum height = 0.5cm, minimum width=0.8cm,thick ]
\tikzstyle{sum}=[circle, fill=blue!10, draw=black,line width=1pt,minimum size = 0.5cm, thick ]
\tikzstyle{ssum}=[circle, fill=blue!10,draw=black,line width=1pt,minimum size = 0.1cm,inner sep=0pt]
\tikzstyle{int1}=[draw, fill=blue!10, minimum height = 0.5cm, minimum width=1cm,thick ]
\tikzstyle{enc}=[draw, fill=blue!10, minimum height = 2.7cm, minimum width=1cm,thick ]
\tikzstyle{int}=[draw, fill=blue!10, minimum height = 1cm, minimum width=1.5cm,thick ]

\title{
Compress-and-Estimate Source Coding\\
for a Vector Gaussian Source
}%
\author{
	\IEEEauthorblockN{Ruiyang Song\IEEEauthorrefmark{1}, Stefano Rini\IEEEauthorrefmark{2},  Alon Kipnis\IEEEauthorrefmark{1}, and  Andrea J. Goldsmith\IEEEauthorrefmark{1} \\}
	
	\IEEEauthorblockA{%
		\IEEEauthorrefmark{1}
		Stanford University, CA, USA \\
		Emails: \texttt{ruiyangs@stanford.edu,
\{kipnisal,andrea\}@wsl.stanford.edu}
	}
	
	\IEEEauthorblockA{%
		\IEEEauthorrefmark{2}
			National Chiao-Tung University, Hsinchu, Taiwan\\
		Email: \texttt{stefano@nctu.edu.tw}
	}
}

\begin{document}
\maketitle

\begin{abstract}
We consider the remote vector source coding problem in which a vector Gaussian source is to be estimated from noisy linear measurements.
For this problem, we derive the performance of the compress-and-estimate (CE) coding scheme and compare it to the optimal performance.
In the CE coding scheme, the remote encoder compresses the noisy source observations so as to minimize the local distortion measure, independent from the joint distribution between the source and the observations.
In reconstruction, the decoder estimates the original source realization from the lossy-compressed noisy observations.
For the CE coding in the Gaussian vector case, we show that, if the code rate is less than a threshold, then the CE coding scheme attains the same performance as the optimal coding scheme.
%
We also introduce lower and upper bounds for the performance gap above this threshold.
In addition, an example with two observations and two sources is studied to illustrate the behavior of the performance gap.
%
%
\end{abstract}

\section{Introduction}

The Distortion-Rate Function (DRF)  describes the minimum attainable average distortion when recovering an information source compressed to within a target bit rate for asymptotically large blocklength.
The setting in which the source sequence is not directly available at the encoder  and only noisy observations are provided is referred to as the \emph{remote} or \emph{indirect} source coding setting \cite[Sec. 3.5]{berger1971rate}.
In general, the optimal coding scheme in the indirect source coding setting depends on the joint statistics of the underlying source and the noisy observations. Therefore, the optimal scheme is not applicable when the encoder is designed without knowledge of this statistics.
Namely, the encoder is either unaware
of the existence of the underlying source and regards the
observed process as the target for compression, or utilizes a general-purpose compression strategy that does not adapt to the underlying source.
%
%
Under these limitations, the encoder will compress its sequence of observations based on a distortion criterion defined only with respect to this observed sequence. 
 The decoder, having full knowledge of the joint statistics of the underlying source and its observations, estimates the source from the output of the encoder.
This compress-and-estimate (CE) scheme was proposed and studied in \cite{KipnisRini2017}.
In this paper we extend the results of \cite{KipnisRini2017} to the Gaussian vector remote source coding problem, thus providing a better understanding of the loss of performance caused by partial system knowledge at the remote encoders.

\subsubsection*{Previous Work}

The indirect source coding problem was first considered in \cite{1057738}, where it was shown that the optimal trade-off between code-rate and distortion is characterized by a single-letter
expression, denoted as the indirect DRF (iDRF).
Moreover, it was proven in \cite{berger1971rate,1056251} that the iDRF can be achieved by first estimating the remote source from the input to the observer, and then encoding this estimate optimally with respect to the code-rate constraint. Therefore, the optimal indirect source coding scheme can be referred to as the estimate-and-compress scheme. The CE scheme for the indirect source coding setting was introduced in \cite{kipnis2016multiterminal} and was motivated by the difficulty in implementing an estimation step before encoding. A single-letter expression characterizing the minimal distortion in the CE setting with multiple encoders was derived in \cite{KipnisRini2017}. Moreover, it was shown in \cite{KipnisRini2017} that this expression has a closed form expression in the quadratic Gaussian setting of a scalar Gaussian source observed through multiple linear measurements corrupted by Gaussian noise. The optimal rate-allocation across the multiple encoders was considered in  \cite{song2016optimal}. \par
Other works that consider communication with non-optimal encoding due to missing source statistics or codebook information include the \emph{oblivious processing} channel coding problem of \cite{4544988}, the minimax source coding of \cite{dembo2003minimax}, and the mismatched encoding problem of \cite{lapidoth1997role}. In particular, as explained in this paper, the CE scheme can be seen as a special case of the mismatched encoding problem.

\subsubsection*{Contributions}
%
%
In this work, we consider the CE scheme in the quadratic Gaussian setting with a vector Gaussian source. Namely, we consider the minimal distortion in estimating a vector Gaussian source from an encoded version of its noisy measurement, where the encoder employs a source code that is optimal with respect to the noisy measurements under a quadratic distortion.
For this setting, we show that the CE scheme is optimal when the code-rate is below a certain threshold that depends on the spectrum of the observed vector. That is, when the code-rate is below this threshold, lack of underlying source statistics causes no penalty in the distortion.
%
%
%
Above this threshold, we derive upper and lower bounds on the distortion difference between the CE scheme and the optimal scheme using an inequality between the arithmetic mean and the geometric mean (AM-GM inequality).
In addition, we consider an example with two sources and two observations to illustrate the behavior of performance gap.
%
%
%
%

\subsubsection*{Paper Organization}
The remainder of the paper is organized as follows: in Sec \ref{sec:problem statement} we introduce the problem formulation.
The main results are presented in Sec. \ref{sec:gaussian vector centralized}.
Sec. \ref{sec:Conclusions} concludes the paper.

Throughout, for the sake of brevity, we only provide a sketch of the proofs in the body text.
The complete proofs are available in the appendices. 

\subsubsection*{Notation}
We denote $[n]=\{1,\ldots,n\}$ for $n\in \Nbb$, and $a^+=\max\{a,0\}$, for $a\in\mathbb R$. Also, we define $\log^+(x)$ as $\log(\max\{1,x\})$.
With $\Iv_{L}$ we indicate the identity matrix of size $L \times L$.
With $\diag(\vv)$ we indicate the matrix with the elements of $\vv$ on the diagonal.
 For a square matrix $\Cv$, denote by $\la_l(\Cv)$ the $l^{\rm th}$ largest eigenvalue. For $\Cv$  positive semi-definite and for $k\in [\rank(\Cv)]$, we define functions $R_k(\Cv)$ and $\theta_k(\Cv,R)$ as follows:
\ea{
R_k(\Cv)=\lcb \p{
0 & k=1 \\
\f{1}{2}\sum_{l=1}^{k} \log\f{\la{_l(\Cv)}}{\la_k(\Cv)}  & 2\leq k \leq \rank(\Cv)  \\
\infty & k= \rank(\Cv)+1,
}
\rnone
\label{eq: R k def}
}
and 	
\ea{
\theta_k(\Cv,R)=2^{-\f{2R}{k}}\lb\prod_{l=1}^{k}\la_l(\Cv)\rb^{\f{1}{k}}.
\label{eq: theta k def}
}
\section{System Model}
\label{sec:problem statement}
\begin{figure}
	\centering
	\begin{tikzpicture}[node distance=2cm,auto,>=latex]
	\node at (-5,0) (source) {$\Xv^n$} ;
	\node [ssum, right of = source,node distance = 1.25 cm](pxy2){$\times$};
	\node [ssum, right of = pxy2,node distance = 1cm](pxy3){$+$};
	\node [above of=pxy2,node distance = 1 cm ] (Av) {$\Av$} ;
	\node [above of=pxy3,node distance = 1 cm ] (Zv) {$\Zv^n$} ;
	\node [int1,right of = pxy3, node distance = 1.5 cm](enc){$\Enc$};
	\node [int1] (dec) [right of=enc, node distance = 3 cm] {$\mathrm{Dec}$};
	\draw[->,line width=1pt] (pxy3) node[above, xshift =0.6 cm, yshift = 0.25 cm] {$\Yv^n$} -- (enc) ;
	\node [right] (dest) [right of=dec, node distance = 1.5cm]{$\widehat{X}^n$};
	%
	\draw[->,line width=1pt] (source) -- (pxy2);
	\draw[->,line width=1pt] (pxy2) -- (pxy3);
	\draw[->,line width=1pt] (Zv) -- (pxy3);
	\draw[->,line width=1pt] (Av) -- (pxy2);
	\draw[->,line width=1pt] (enc) -- node[above, xshift = 0.25cm, yshift = 0.25cm, text width  = 2.5 cm] {\scriptsize $m \in \lcb 1 \ldots 2^{ \lfloor n R \rfloor} \rcb$} (dec);
	%
	\draw[->,line width=1pt] (dec) -- (dest);
	\end{tikzpicture}
	\caption{
		Vector Gaussian remote source coding problem with  noisy linear measurement.
	}
	\label{fig:remote_source_coding_system_model}
	\vspace{-.3 cm}
\end{figure}
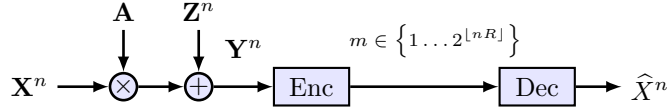

We consider the indirect source coding setting in Fig. \ref{fig:remote_source_coding_system_model} in which an $M$-dimensional Gaussian source is observed through a vector additive white Gaussian noise (AWGN) channel at a remote encoder.
%
The random source sequence $\Xv^n =(X_1^n \ldots X_M^n )$  is obtained through $n$ i.i.d. draws from the jointly Gaussian distribution with zero mean and covariance matrix $\Iv$.
%
The remote encoder obtains the noisy observation vector  $\Yv=(Y_1^n \ldots Y_L^n )$ with
\ea{
	\Yv^n=\Av \Xv^n+\Zv^n,
	\label{eq:observation model}
}
for some $\Av\in \mathbb R^{L\times M}$, assumed full rank, with noise $\Zv \sim \mathcal N(\bm 0,\sigma^2 \Iv_{L})\in \mathbb R^L$.
%
%
The encoder produces the index $m \in \left\{1 \ldots 2^{ \lfloor n R \rfloor} \right\}$ to encode its observation and the index $m$ is noiselessly communicated to a central processing unit that produces the reconstruction sequence $\widehat{\bf X}^n(m)$. \par
Given a value $R$, we wish to determine the minimum average quadratic distortion between the original source sequence $\Xv^n$ and its reconstruction $\widehat{\bf X}^n$, normalized over the source dimension $M$, which is:
\ea{
	\mathbb E d(\Xv^n,\widehat{\bf X}^n)
	&  \triangleq \f{1}{M} \f{1}{n} \sum_{i=1}^n \mathbb \Ebb \lsb
	\left\Vert\Xv_i-\widehat{\bf X}_i(m)\right\Vert^2
	\rsb,
	\label{eq:distortion_def_1}
}
where the expectation is taken with respect to all source and  channel realizations.

\begin{rem}
The model in \eqref{eq:observation model} in which $\Xv$ has a general covariance matrix $\bm{\Sigma_X}$ can be reduced to the case $\bm{\Sigma_X}=\Iv$ without loss of generality.
This is obtained by letting $\Xv'=\bm{\Sigma_{X}}^{-1/2}\Xv$ and $\Av'=\Av\bm{\Sigma_{X}}^{1/2}$. Hence $\Yv^n=\Av' \Xv'^n+\Zv^n$, where the whitened source vector $\Xv'\sim\mathcal N({\bf 0},\Iv)$.
\end{rem}
In the following, we assume that the matrix $\Av$ in \eqref{eq:observation model} is full rank and define $r=\min\{M,L\}$. We use $\{\la_l, \ l\in [L]\}$ to denote the eigenvalues of $\Av\Av^T$, sorted in descending order.
Note that the covariance matrix of $\Yv$, i.e. $\bm{\Sigma_Y}$, and the conditional covariance matrix of $\Xv|\Yv$, i.e. $\bm{\Sigma_{X|Y}}$, are respectively obtained as
\eas{
\bm{\Sigma_Y} & =\Av\bm {\Sigma_X}\Av^T+\sigma^2\Iv
\label{def:Lambda_Y}, \\
\bm{\Sigma_{X|Y}} &=\bm{\Sigma_X} \Av^T(\Av\bm{\Sigma_X} \Av^T+\bm{\Sigma_W})^{-1}\Av\bm {\Sigma_X}.
}{\label{def:Lambda_Y all}}
Following \eqref{def:Lambda_Y all},  we have that  $\{\la_l/(\la_l+\sigma^2), \   l\in [r] \}$ are the eigenvalues
of  $\bm{\Sigma_{X|Y}}$ and $\{\la_l+\sigma^2, \  l\in [r] \}$ are the eigenvalues
of  $\bm{\Sigma_{Y}}$, both in descending order as per $\{\la_l,  \   l\in [r] \}$.

\subsubsection*{Indirect Source Coding}
The minimal distortion in the indirect source coding setting of Fig. \ref{fig:remote_source_coding_system_model} is described by the (information)
iDRF \cite[p.78-81]{berger1971rate}:
\ea{ \label{eq:indirect_coding_thm}
D_{X|Y}(R) = \inf \mathbb E d(\Xv,\widehat{\Xv}),
}
where the infimum is taken over all joint probability distributions of $\Yv$ and $\widehat{\Xv}$ such that the per letter mutual information
$I(\Yv; \widehat{\Xv})$ does not exceed $R$.
%

%
%

The iDRF for the model in \eqref{eq:observation model} can be rewritten as
\ea{
\label{eq:idrf_Gaussian}
D_{X|Y}(R)&  = \f 1 M \trace \lb \mmse(\Xv|\Yv) \rnone \\
& \quad \quad  + \lnone \min_{\widehat{\Xv}: I(\Yv,\widehat{{\Xv}})\leq R} \Ebb_{\Yv} \lsb || \Ebb[ \Xv|\Yv]-\widehat{\Xv} ||^2 \rsb\rb, \nonumber
}
where $\mmse(\Xv|\Yv)$ is the minimum mean square error (MMSE) when estimating $\Xv$ from $\Yv$, i.e.
\eas{
\Ebb[ \Xv|\Yv]& =\Av^T(\Av \Av ^T+\sgs \Iv)^{-1}\Yv, \\
\mmse(\Xv|\Yv)& =\lb \f 1 \sgs \Av \Av^T+\Iv\rb^{-1}.
}
The solution of the optimization in \eqref{eq:indirect_coding_thm}  through the formulation in \eqref{eq:idrf_Gaussian} results in the classic water-filling assignment of the
compression rate for each observation.
This solution yields a simple expression for the inverse of the iDRF, which is the rate-distortion function
\ea{
R_{X|Y}(D)=\f 1  M \sum_{l=1}^L \f { \la_l}{\la_l+\sgs} +  \sum_{l=1}^L \f 12 \log \f 1 {D_l}\f { \la_l}{\la_l+\sgs},
\label{eq:R idrf}
}
where
\ea{
D_l= \lcb  \p{
\theta &   \f { \la_l}{\la_l+\sgs} > \theta \\
\f { \la_l}{\la_l+\sgs} & \f { \la_l}{\la_l+\sgs}\leq \theta,
}
\rnone
\label{eq:D idrf}
}
where $\theta$ is chosen so that $\sum_l D_l=D$.
The expression in \eqref{eq:D idrf} shows that, in the optimal compression scheme, $\Ebb[(\Xh_l-X_l)^2|\Yv]=D_l$  for $D_l$ in \eqref{eq:D idrf}.
In other words, the MMSE error in estimating each source component is controlled by the solution of the  water-filling problem so that if the $i^{\rm th}$ eigenvalue is below the water level, no rate is assigned to the compression of the estimate of the $i^{\rm th}$ observation.

\subsubsection*{Compress-and-Estimate Vector Source Coding}
%
The CE setting \cite{kipnis2016multiterminal} considers the remote source coding problem
in which each remote encoder compresses its noisy observation sequence so as to minimize a local distortion measure that depends
only on the distribution of its observed sequence, and is otherwise independent from the distribution of the underlying
source.
The CE-DRF is the single letter expression of the distortion that can be
attained by the CE coding scheme for the case of an i.i.d. source observed through a memoryless channel.
Given a local distortion measure $d_l$ and a probability distribution $P^*_{\widehat{\Yv},\Yv}$ that satisfies $I(\Yv;\widehat{\Yv}) = R$ and $\mathbb E d(Y_l,\widehat{Y}_l) = D(R)$, where $D(R)$ is the classic distortion-rate function with respect to compressing  $\Yv$ under distortion measure $d_l$, the CE-DRF is defined as
\ea{
\label{eq:ce-drf-def}
D_{CE}(R) =  \inf D \lb\Xv,\widehat{\bf X}(\widehat{\Yv})\rb,
}
where the infimum is over all estimators of $\Xv$ given the noisy reconstructions $\widehat{\Yv}$.
%
For the observation model in \eqref{eq:observation model}, the quadratic distortion in \eqref{eq:distortion_def_1} at the central unit and as a local distortion at the remote encoder, the CE-DRF is expressed as
\ea{
D_{CE}(R) =
 \Ebb \lsb ||\Xv-E[\Xv|\widehat{\Yv}] ||^2 \rsb,
\label{eq:CE_Gaussian}
}
where
%
%
the joint distribution $P_{\Yv,\widehat{\Yv}}$
is described by the backward Gaussian channel $\Yv= \widehat{\Yv}+\Uv\Zv$, where $\Uv$ is an orthogonal matrix that diagonalizes $\bm{\Sigma_Y}$, and the noise $\Zv \sim \Ncal({\bf 0},\bm{\Sigma_Z})$ with covariance matrix $\bm{\Sigma_Z}=\diag\{\sigma_{Z_1}^2,\ldots,\sigma_{Z_L}^2\}$. Here, $\sigma_{Z_l}^2=\min(\la_l+\sigma^2,\theta)=D_l$.
%

\section{Main Result}
\label{sec:gaussian vector centralized}
In the following, we derive conditions under which the CE-DRF in the vector quadratic Gaussian setting equals the iDRF. In addition,
%
%
we derive bounds, both upper and lower, to the performance gap between the CE-DRF and the iDRF.
%
%
We also provide a two-dimensional example and characterize the performance gap in different regions of $R$.

\subsection{Conditions for Equality}
\label{sec:Conditions for equality}
We begin by deriving the conditions under which the CE-DRF and the iDRF coincide.
To do so, we re-write the iDRF as the inverse of the rate-distortion  function in \eqref{eq:R idrf}.
\begin{prop}\label{prop:idrf linear}
When $R$ satisfies
\ea{
R_{k}(\bm{\Sigma_{X|Y}}) < R \leq R_{k+1}(\bm{\Sigma_{X|Y}}),
\label{eq:idrf linear interval}
}
for  $R_k(\bm\Sigma_{X|Y})$ in \eqref{eq: R k def}  and $k \in [r]$, the iDRF in \eqref{eq:idrf_Gaussian} is obtained as
\ea{
D_{X|Y}(R)= 1-\f{1}{M}\sum_{l=1}^{k}{\f{\la_l}{\la_l+\sigma^2}}+\f{k}{M} \theta_k(\bm{\Sigma_{X|Y}},R),
\label{eq:idrf linear}
}
for $\theta_k(\bm{\Sigma_{X|Y}},R)$ in \eqref{eq: theta k def}.
\end{prop}

In Prop. \ref{prop:idrf linear}, we express $D_{X|Y}(R)$ as a piecewise function of $R$ over the intervals
 $\lcb (R_{k}(\bm{\Sigma_{X|Y}}), R_{k+1}(\bm{\Sigma_{X|Y}})] \rcb_{k=1}^{r+1}$ with $\cup_{k=1}^r(R_{k}(\bm{\Sigma_{X|Y}}), R_{k+1}(\bm{\Sigma_{X|Y}})]=\Rbb^+$.
In each one of those intervals, the water-filling solution prescribes for $k$ eigendirections of $\bm{\Sigma_{X|Y}}$ to be compressed at the remote encoder.
In other words, $k$ in \eqref{eq:idrf linear} corresponds to the number of $D_l>\theta$ in \eqref{eq:R idrf}.

We then obtain an expression of the CE-DRF.
%
%

\begin{prop}\label{prop:CE centralized}
When the rate $R$ satisfies
\ea{
R_{k}(\bm{\Sigma_{Y}}) < R \leq R_{k+1}(\bm{\Sigma_{Y}}),
\label{eq:cedrf_central interval}
}
the CE-DRF in \eqref{eq:CE_Gaussian} is obtained as
\ea{
D_{CE}(R) & =   
1-\f 1 M \sum_{l=1}^{k} \f {\la_l} {\la_l+\sgs}  \nonumber \\
& \quad \quad +\f{\theta_k(\bm{\Sigma_Y},R)}{M} \sum_{l=1}^k \f {\la_l}{(\la_l+\sgs)^2}.
\label{eq:cedrf_central}
	}
\end{prop}
%
%
%
By comparing the expressions in Prop. \ref{prop:idrf linear} and Prop. \ref{prop:CE centralized} we
note that, for
$
R \leq \min \{ R_2( \bm{\Sigma_Y}), R_2( \bm{\Sigma_{X|Y}}) \},
$
the expressions in \eqref{eq:idrf linear} and \eqref{eq:cedrf_central} are equal.
%
%
This condition for equality can be generalized as in the following proposition.

\begin{prop}
\label{prop:equality small rate}
Define
\ea{
r_0
=\max_k 
\left\{\la_k=\frac{1-2c(\la_1)\sigma^2\pm\sqrt{1-4c(\la_1)\sigma^2}}{2c(\la_1)}\right\},
\label{def:r0}
}
where $c(\la_1)={\la_1}/{(\la_1+\sigma^2)^2}$.
Or equivalently, $r_0$ is defined as the integer such that  $\la_k/(\la_k+\sigma^2)^2=\la_1/(\la_1+\sigma^2)^2$ if and only if $k\leq r_0$.
Then, if  $R\leq \min\{ R_{r_0+1}( \bm{\Sigma_Y}), R_{r_0+1}( \bm{\Sigma_{X|Y}})
\}$, the expressions in \eqref{eq:idrf linear} and \eqref{eq:cedrf_central} are equal.
\end{prop}
%
The result in Prof. \ref{prop:equality small rate} generalizes a previous result in \cite{kipnis2016multiterminal} where equality was shown
for the case of $L=M=1$.
Note that if $r_0=L=M$, then we have $D_{CE}(R)=D_{X|Y}(R)$ for any $R$.
The functions $D_{X|Y}(R)$ and $D_{CE}(R)$ are continuous decreasing functions of $R$. We note that $D_{CE}(R)$ is only smooth between each pair of two rate thresholds, that is in the intervals $[R_k(\bm{\Sigma_{Y}}),R_{k+1}(\bm{\Sigma_{Y}})]$. Each rate threshold $R_k(\bm{\Sigma_{Y}})$ determines a change in the slope of $D_{CE}(R)$. %
Despite the fundamental difference between these two functions, in the low rate regime of Prop.  \ref{prop:equality small rate}
the two functions actually coincide.

\subsubsection*{Discussion} Note that both the CE and the optimal compression schemes suffer from a distortion $\mmse(\Xv|\Yv)$, which  follows from the fact that $\Yv$, instead of $\Xv$, is observed. Moreover, the forward channel between the observation and its reconstruction for both transmission schemes corresponds to a vector additive Gaussian noise channel.
%
%
In the optimal compression scheme, the noise is added along the eigenvectors of the covariance matrix $\bm{\Sigma_{X|Y}}$, while in the CE scheme the noise follows the direction of the eigenvectors of the matrix $ \bm{\Sigma_{\Yv}}$.
This distinction implies that the CE scheme might allocate compression resources to eigendirections which are not useful in estimating the underlying source $\Xv$.
As an example, assume  that there exists $\la_l=0$.
Recall \eqref{def:Lambda_Y},  
when the $l^{\rm th}$ eigenvalue of $\Av\bm {\Sigmav_X}\Av^T$ is zero, the $l^{\rm th}$ eigenvalue of $\bm{\Sigma_Y}$ equals $\sigma^2$, which solely comes from the covariance matrix of noise $\sigma^2\Iv$, and the $l^{\rm th}$ eigenvalue of $\bm{\Sigma_{X|Y}}$ satisfies
${\la}_l(\bm{\Sigma_{X|Y}})=\la_l/(\la_l+\sigma^2)=0$.
Hence, the $l^{\rm th}$ components in both optimal and CE settings are pure noise and contain no information of the source.
 For the optimal scheme, we see that the compression rate allocated for this component is always zero.
In other words, the optimal scheme never activates a component that does not contain source information.
However, we note that the CE scheme might activate components that are pure noise when $R$ is sufficiently large.
This is due to the fact that, in the optimal scheme, the encoder knows the joint statistics and can therefore avoid wasting rate resources on useless observations. In contrast, the encoder in the CE setting cannot recognize a pure noise component since it lacks knowledge of the source.
On the other hand, when the rate is sufficiently small, only the largest eigendirection is actively compressed, in which case, perhaps surprisingly, the CE performance equals the optimal performance.
\subsection{Performance Gap}
We next upper bound the performance gap between the CE and the optimal performance, defined as
\ea{
G(R)= D_{CE}(R)-D_{X|Y}(R).
\label{eq:gap def}
}
 for the regimes in which the equality conditions of Prop. \ref{prop:equality small rate} does not hold.
\begin{thm}
\label{thm:performance gap}
The difference between the CE-DRF and the iDRF is bounded as
\ea{
G(R)
  \leq \f{L}{M} \f{\la_1+\sigma^2}{4\sigma^2} 2^{-2 \f R L }.
\label{eq:gap bound 1}
}	
%
\end{thm}
\begin{IEEEproof}
Only a sketch of the proof is presented here.
As an example, assume that there exists an $R$ for which \eqref{eq:idrf linear interval} and \eqref{eq:cedrf_central interval} hold for the same  $k$.
As argued above, this implies that the two schemes actively compress the same number of components.
For this value of $R$ we have
{\small
\ea{
& M\lb 1-D_{CE}(R)\rb
\nonumber \\
&=\sum_{l=1}^{k}\f{\la_l}{\la_l+\sigma^2}+\f{2^{-\f{2R}{k}}}{M}\lb\prod_{l=1}^{k}(\la_l+\sigma^2)\rb^{\f{1}{k}}k
\lb \sum_{l=1}^{k} \f 1 k \f{\la_l}{(\la_l+\sigma^2)^2} \rb
\nonumber\\
& \leq \sum_{l=1}^{k}\f{\la_l}{\la_l+\sigma^2}+\f{2^{-\f{2R}{k}}}{M}\lb\prod_{l=1}^{k}(\la_l+\sigma^2)\rb^{\f{1}{k}}k
\lb \prod_{l=1}^{k} \f{\la_l}{(\la_l+\sigma^2)^2} \rb^{1/k}
\label{eq:dce>=idrfx}  \\
%
%
& \leq M(1- D_{X|Y}(R)), \nonumber
}
}
where \eqref{eq:dce>=idrfx}  follows from the classic AM-GM inequality.
Through a similar reasoning, the inequality in \eqref{eq:gap bound 1} is obtained by  exploiting a reverse AM-GM inequality in \cite{nguyen2008reversing} to bound the
largest difference between the CE and the optimal performance.
\end{IEEEproof}
%
Th. \ref{thm:performance gap} shows that
the difference between the CE-DRF and the optimal distortion is upper bounded by a function that decreases exponentially with $R/L$. 
To complement the result in Th. \ref{thm:performance gap}, we introduce the following lower bound to the performance gap between the CE and the optimal performance.
\begin{thm}
\label{thm:performance gap 2}
For $R>R_2(\bm{\Sigma_Y})$, the difference between CE-DRF and iDRF is lower bounded as
\begin{align}
%
G(R) \geq  \frac{\la_{L}+\sigma^2}{M}\left(\frac{\sqrt{\la_1}}{\la_1+\sigma^2}-\frac{\sqrt{\la_2}}{\la_2+\sigma^2}\right)^22^{-\frac{2R}{L}}.
\label{eq:performance gap 2}
\end{align}	
\end{thm}
\begin{IEEEproof}
We provide a sketch of proof here. Suppose $k$ satisfies \eqref{eq:cedrf_central interval}. We have
{\small
\begin{align}
&~~~D_{CE}(R)-D_{X|Y}(R)\nonumber\\
&\geq \frac{2^{-\frac{2R}{k}}}{M}\prod_{l=1}^k (\la_l+\sigma^2)^{\frac{1}{k}}\left(\sum_{l=1}^k\frac{\la_l}{(\la_l+\sigma^2)^2}+  \rnone \nonumber \\
& \quad \quad \quad \lnone-k\left(\prod_{l=1}^k\frac{\la_l}{(\la_l+\sigma^2)^2}\right)^{\frac{1}{k}}\right)\nonumber\\	
&\geq
\frac{2^{-\frac{2R}{k}}}{M}\prod_{l=1}^k (\la_l+\sigma^2)^{\frac{1}{k}}\left(\max_{l\leq k}\frac{\sqrt{\la_l}}{(\la_l+\sigma^2)}-\min_{l\leq k}\frac{\sqrt{\la_l}}{(\la_l+\sigma^2)}\right)^2.
\end{align}	
}
The first inequality follows from the fact that $k$ satisfies \eqref{eq:cedrf_central interval} but not necessarily \eqref{eq:idrf linear interval}, and hence could be
a non-optimal choice for the setting with full knowledge. The second inequality follows from a lower bound on the difference between AM and GM in \cite[Sec. II]{tung1975lower}.
Finally, \eqref{eq:performance gap 2} follows by noting the bound on $R$ given in \eqref{eq:cedrf_central interval} and the fact that $\la_l$'s are in descending order.
\end{IEEEproof}

Recall that the performance gap can be zero for small $R$,
the lower bound \eqref{eq:performance gap 2} is valid for $R>R_2(\bm{\Sigma_Y})$.
%
For $R\leq R_2(\bm{\Sigma_Y})$, the obvious lower bound $D_{CE}(R)-D_{X|Y}(R)\geq 0$ could be tight.

\smallskip

The results in Th. \ref{thm:performance gap} and Th. \ref{thm:performance gap 2} show that the gap between the CE and the optimal performance decays exponentially
in the rate-per-observation, $R/L$.
The upper and lower bounds are always monotonically decreasing but, from numerical observations, they appear to be loose in the region for small $R$.
Nonetheless they correctly capture the asymptotic decrease in the gap between the performance of the two schemes.

\subsection{Two Observations, Two Sources Example}

In this section we consider the case with two sources and two observations, i.e. $M=L=2$.
For brevity, we consider only the scenario \ea{
	\la_1/(\la_1+\sigma^2)^2\leq \la_2/(\la_2+\sigma^2)^2.
\label{eq:condition_2}
}
Recall that, by the definition in \eqref{eq: R k def}, $R_1(\bm{\Sigma_{X|Y}})=R_1(\bm{\Sigma_Y})=0$, additionally
\ea{
 R_2(\bm{\Sigma_Y})=\frac{1}{2}\log\frac{\la_1+\sigma^2}{\la_2+\sigma^2} \nonumber \\
 \leq \frac{1}{2}\log\frac{\la_1/(\la_1+\sigma^2)}{\la_2/(\la_2+\sigma^2)}=R_2(\bm{\Sigma_{X|Y}}).
}
Following the assumption in \eqref{eq:condition_2}.
Given the considerations above, we derive the performance gap in three regions:
(i) $R \in (0,R_2(\bm{\Sigma_{X|Y}})]$,
(ii) $R \in (R_2(\bm{\Sigma_{X|Y}}), R_2(\bm{\Sigma_Y})]$, and (iii) $R \in (R_2(\bm{\Sigma_Y}), \infty)$.

\noindent
$\bullet$ \underline{$R \in (0,R_2(\bm{\Sigma_{X|Y}})]$:} This is the region of equality in Prop. \ref{prop:equality small rate}.

\noindent
$\bullet$ \underline{$R \in (R_2(\bm{\Sigma_{X|Y}}), R_2(\bm{\Sigma_Y})]$:} Here we have
 \begin{align}
%
G(R)	 =\frac{1}{2}\left(\sqrt{\frac{\la_1}{\la_1+\sigma^2}}2^{-R}-\sqrt{\frac{\la_2}{\la_2+\sigma^2}}\right)^2.
	 \end{align}
	 In this region, the performance gap increases with $R$, and the maximal gap is
	 \ea{
	 	\max_R G(R)
	 	 =\frac{1}{2}(\la_2+\sigma^2)\left(\frac{\sqrt{\la_1}}{(\la_1+\sigma^2)}-\frac{\sqrt{\la_2}}{(\la_2+\sigma^2)}\right)^2,
	 }
	 which is achieved at $R=R_2(\bm{\Sigma_Y})$.

\noindent
$\bullet$ \underline{$R \in (R_2(\bm{\Sigma_Y}), \infty)$:}
In this case we have
\ea{
G(R)
 &= 2^{-R}\sqrt{(\la_1+\sigma^2)(\la_2+\sigma^2)}\nonumber\\
	 &\cdot\left(\frac{1}{2}\sum_{l=1}^2\frac{\la_l}{\la_l+\sigma^2}-\left(\prod_{l=1}^2\frac{\la_l}{\la_l+\sigma^2}\right)^{\frac{1}{2}}\right).
}
We see that in this region, the performance gap decays exponentially in $2R/L=R$, which corresponds to the behavior predicted by Th. \ref{thm:performance gap} and Th. \ref{thm:performance gap 2}.
%
%
%
%
%
%

\smallskip
\begin{figure}
	\centering
	\includegraphics[width=0.5\textwidth]{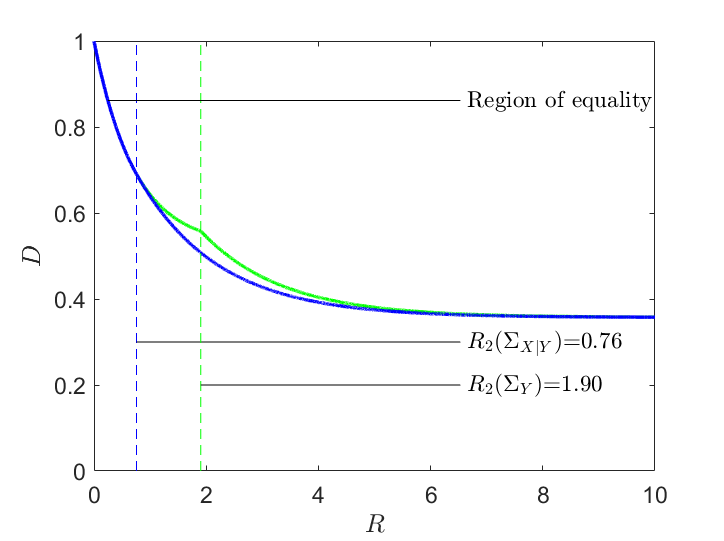}
	\caption{Comparison between $D_{CE}$ (green) and $D_{X|Y}$ (blue).}
	\label{fig:comp_not_equal}
\end{figure}
\begin{figure}
	\centering
	\includegraphics[width=0.5\textwidth]{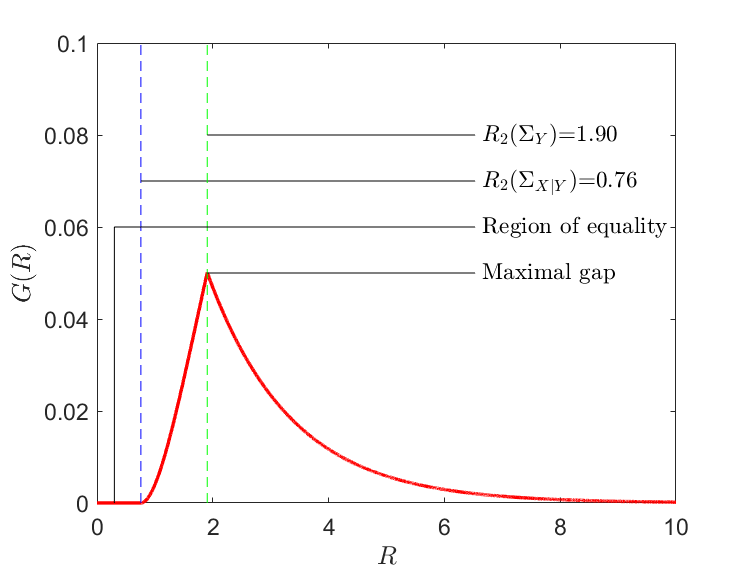}	
	\caption{Performance gap $G(R)=D_{CE}(R)-D_{X|Y}(R)$.}
	\label{fig:gap}
	\vspace{-0.5 cm}
\end{figure}
Let us introduce a numerical evaluation for the case  $L=M=2$ with $\la_1=20$, $\la_2=0.5$ and $\sigma^2=1$, yielding $R_2(\bm{\Sigma_{X|Y}})=0.76<R_2(\bm{\Sigma_Y})=1.90$.
In Fig. \ref{fig:comp_not_equal}, we  plot the CE-DRF together with the iDRF, which are monotonically decreasing in $R$.
The function $D_{CE}(R)$ is smooth on intervals $[0,R_2(\bm{\Sigma_Y})]$ and $(R_2(\bm{\Sigma_Y}),+\infty)$,
 while the derivative is discontinuous at the point $R=R_2(\bm{\Sigma_Y})$.

Fig. \ref{fig:gap} focuses on the performance gap, $G(R)$ in \eqref{eq:gap def}: when $0<R\leq R_2(\bm{\Sigma_{X|Y}})$, we have $D_{CE}(R)=D_{X|Y}(R)$;
 when $R_2(\bm{\Sigma_{X|Y}})<R\leq R_2(\bm{\Sigma_{X|Y}})$, the performance gap is increasing.
 The performance gap starts to decrease at $R=R_2(\bm{\Sigma_Y})$ so that  the maximal gap is $G(R)=0.05$, which is achieved at $R=R_2(\bm{\Sigma_Y})=1.90$.

\section{Conclusion}
\label{sec:Conclusions}
We have derived the performance of compress-and-estimate (CE) coding for a  Gaussian vector, observed at the remote encoder through linear observations and further
corrupted by additive white Gaussian noise.
In the CE coding scheme, the remote encoder compresses its observation according to a local distortion measure which depends solely on the observation distribution
and rate-per-symbol constraint.
 An estimator receives the encoded observations and uses them to estimate the remote source sequence.
%
For this setting, we showed that when the rate is smaller than a threshold, the CE setting attains the optimal source coding performance.
This threshold is obtained as a rather straightforward  function of the eigenvalues of the observation matrix.
Since the operation at the remote encoder depends only on the distribution of the observation,  this result shows instances in which the optimal coding performance
can be attained without the full system knowledge at the remote encoder.
In addition, we derived upper and lower bounds on the performance gap between the CE scheme and the optimal scheme where the encoder has full knowledge of the underlying source statistics, which shows that the decay in the performance loss is exponential in the rate-per-observation in the region where the rate-per-symbol is large.
%
%
%
Finally, for the case of two observations and two sources, a complete characterization of the behavior of the performance gap is derived.

\bibliographystyle{IEEEtran}
\bibliography{mismatched}

\begin{thebibliography}{10}
\providecommand{\url}[1]{#1}
\csname url@samestyle\endcsname
\providecommand{\newblock}{\relax}
\providecommand{\bibinfo}[2]{#2}
\providecommand{\BIBentrySTDinterwordspacing}{\spaceskip=0pt\relax}
\providecommand{\BIBentryALTinterwordstretchfactor}{4}
\providecommand{\BIBentryALTinterwordspacing}{\spaceskip=\fontdimen2\font plus
\BIBentryALTinterwordstretchfactor\fontdimen3\font minus
  \fontdimen4\font\relax}
\providecommand{\BIBforeignlanguage}[2]{{%
\expandafter\ifx\csname l@#1\endcsname\relax
\typeout{** WARNING: IEEEtran.bst: No hyphenation pattern has been}%
\typeout{** loaded for the language `#1'. Using the pattern for}%
\typeout{** the default language instead.}%
\else
\language=\csname l@#1\endcsname
\fi
#2}}
\providecommand{\BIBdecl}{\relax}
\BIBdecl

\bibitem{berger1971rate}
T.~Berger, \emph{Rate-distortion theory: A mathematical basis for data
  compression}.\hskip 1em plus 0.5em minus 0.4em\relax Englewood Cliffs, NJ:
  Prentice-Hall, 1971.

\bibitem{KipnisRini2017}
\BIBentryALTinterwordspacing
A.~Kipnis, S.~Rini, and A.~J. Goldsmith, ``Compress and estimate in
  multiterminal source coding,'' 2017, unpublished. [Online]. Available:
  \url{https://arxiv.org/abs/1602.02201}
\BIBentrySTDinterwordspacing

\bibitem{1057738}
R.~Dobrushin and B.~Tsybakov, ``Information transmission with additional
  noise,'' \emph{IRE Transactions on Information Theory}, vol.~5, no.~8, pp.
  293--304, 1962.

\bibitem{1056251}
H.~Witsenhausen, ``Indirect rate distortion problems,'' \emph{Information
  Theory, IEEE Transactions on}, vol.~26, no.~5, pp. 518--521, Sep 1980.

\bibitem{kipnis2016multiterminal}
A.~Kipnis, S.~Rini, and A.~J. Goldsmith, ``Multiterminal compress-and-estimate
  source coding,'' in \emph{Information Theory (ISIT), IEEE International
  Symposium on}, 2016, pp. 540--544.

\bibitem{song2016optimal}
R.~Song, S.~Rini, A.~Kipnis, and A.~J. Goldsmith, ``Optimal rate allocation in
  multiterminal compress-and-estimate source coding,'' in \emph{Information
  Theory Workshop (ITW), IEEE}, 2016, pp. 111--115.

\bibitem{4544988}
A.~Sanderovich, S.~Shamai, Y.~Steinberg, and G.~Kramer, ``Communication via
  decentralized processing,'' \emph{Information Theory, IEEE Transactions on},
  vol.~54, no.~7, pp. 3008--3023, July 2008.

\bibitem{dembo2003minimax}
A.~Dembo and T.~Weissman, ``The minimax distortion redundancy in noisy source
  coding,'' \emph{Information Theory, IEEE Transactions on}, vol.~49, no.~11,
  pp. 3020--3030, 2003.

\bibitem{lapidoth1997role}
A.~Lapidoth, ``On the role of mismatch in rate distortion theory,''
  \emph{Information Theory, IEEE Transactions on}, vol.~43, no.~1, pp. 38--47,
  1997.

\bibitem{nguyen2008reversing}
T.~L. Nguyen, ``Reversing the arithmetic mean--geometric mean inequality,''
  \emph{Research report collection}, vol.~11, no. Supp, 2008.

\bibitem{tung1975lower}
S.~Tung, ``On lower and upper bounds of the difference between the arithmetic
  and the geometric mean,'' \emph{Mathematics of Computation}, vol.~29, no.
  131, pp. 834--836, 1975.

\end{thebibliography}

\onecolumn
\appendices
\section{Proof of Proposition \ref{prop:idrf linear}}
\label{app:idrf linear}

The expression in \eqref{eq:idrf linear} is substantially a convenient formulation of the inverse of \eqref{eq:R idrf} using the structure of the  water filling solution.
%
In optimal coding scheme, the encoder first estimates the source $\Xv$ based on the observation vector $\Yv$ and produces the MMSE source estimate $\Xv|\Yv$.
%
%
Recall that $\bm{\Sigma_X}=\Iv$ so that the covariance matrix of the source estimate, denote as $\cov(X|Y)=\Sigma_{X|Y}$, can be expanded as
\begin{align}
\bm{\Sigma_{X|Y}}
&=\bm{\Sigma_X}\Av^T(\Av\bm{\Sigma_X}\Av^T+\bm{\Sigma_W})^{-1}\Av\bm{\Sigma_X}\nonumber\\
&=\Av^T(\Av\Av^T+\sigma^2\Iv)^{-1}\Av \\
&=\lb \f {\Av\Av^T} {\sgs}+\Iv\rb^{-1}\Av,
\end{align}
from which we conclude that the eigenvalues of $\bm{\Sigma_{X|Y}}$ are
\begin{equation}
\la_l(\bm{\Sigma_{X|Y}})=\f {\la_l} {\la_l+\sigma^2},
\end{equation}
for $l\in[M]$.
In the optimal rate allocation of \eqref{eq:R idrf}, the $l^{\rm th}$ observation is compressed with rate 
\begin{equation}
R_l(\theta)=\f{1}{2}\log^+\lb\f{\la_l(\bm{\Sigma_{X|Y}})}{\theta}\rb,
\label{eq:rates opt}
\end{equation}
where $\theta$ is determined by the sum rate constraint 
\ea{
\sum_{l=1}^{M}R_l(\theta)=R.
\label{eq:sol theta}
}
The function $\theta_{k}(\bm{\Sigma_{X|Y}},R)$ in \eqref{eq: theta k def} represents the solution of  \eqref{eq:sol theta}  in $\theta$ when $k$ rates in \eqref{eq:rates opt}
are strictly positive. 

Next, let $k_I$ be the index $k$ that satisfies the inequality \eqref{eq:idrf linear interval}, then the normalized average quadratic distortion in terms of the remote source $\Xv$ is given by
\begin{align}
D(\theta(R))
&=\f{1}{M}\lb{\rm Tr}(\bm{\Sigma_X})-\sum_{l=1}^{M}\lb\la_l(\bm{\Sigma_{X|Y}})-\theta_{k_I}(\bm{\Sigma_{X|Y}},R)\rb^+\rb\nonumber\\
&=1-\f{1}{M}\sum_{l=1}^{k_I}\f{\la_l}{\la_l+\sigma^2}+\f{k_I}{M}\theta_{k_I}(\bm{\Sigma_{X|Y}},R),
\end{align}
which is the desired result.
\section{Proof of Proposition \ref{prop:CE centralized}}
\label{app:CE centralized}
Unlike in the optimal scheme, in the compress-and-estimate scheme the encoder compresses its observations according to the distribution $P_\Yv$ and a local distortion measure.
Upon reconstruction, the decoder obtains the lossy-compressed observation $\widehat{\Yv}$ minimizes the quadratic distortion between $\widehat{\Yv}$ and $\Yv$ 
as in \eqref{eq:distortion_def_1} for the given rate constraint.
%
The joint distribution $P_{\Yv,\widehat{\Yv}}$ which minimizes the quadratic distortion is  easily expressed through the backward channel formulation as
\ea{
\Yv= \widehat{\Yv}+\Uv\Zv,
}
where $\Uv$ is an orthogonal matrix that diagonalizes $\bm{\Sigma_Y}$, $\Zv \sim \Ncal({\bf 0},\bm{\Sigma_Z})$ with $\bm{\Sigma_Z}=\diag([\sigma_{Z_1}^2,\ldots,\sigma_{Z_L}^2])$
and for $\sigma_{Z_l}^2=\min(\la_l+\sigma^2,\theta)=D_l$.
In other words, the term $D_l$ is the variance of the additive noise corrupting the $l^{\rm th}$ observation while $\theta$ is chosen so that the sum rate 
constraint 
\ea{
\sum_{l=1}^{M}R_l(\theta)=R,
\label{eq:sol theta 2}
}
is met with equality, in which case
\ea{
	R_l(\theta)=\f{1}{2}\log^+\lb\f{\la_l}{\theta}\rb.
\label{eq:rate opt 2}
}
As in App. \ref{app:idrf linear}, the function  $\theta_{k}(\bm{\Sigma_{Y}},R)$ in \eqref{eq: theta k def} represents the solution of  \eqref{eq:sol theta 2}
in $\theta$ when $k$ rates in \eqref{eq:rate opt 2} are strictly positive. 
The joint distribution between $\Xv$ and $\widehat{\Yv}$ can be expressed as 
\ea{
	\widehat{\Yv}=\Pv\Xv+\bm \eta,
}
where $\Pv=\Jv\Uv^T \bm A$, and $\bm \eta=\Jv\Uv^T\Wv+\Jv^{1/2}\Dv^{1/2}\Nv$
with $\Jv$ defined as
\begin{equation}\label{def:J}
\Jv =\diag([1-2^{-2R_1},\ldots,1-2^{-2R_L}]),
\end{equation}
while $\Dv$ defined as
\ea{
	\Dv=\diag([D_1,\ldots,D_L]),
	\label{def:D}
}
and $\Nv\sim\mathcal N({\bf 0}, \Iv)$. The covariance matrix of the noise vector $\bm{\eta}$ is, instead, obtained as $\bm{\Sigma_\eta}=\sigma^2\Jv^2+\Jv\Dv$.

The CE-DRF is the normalized minimum mean square error of estimating $\Xv$ from $\widehat{\Yv}$, accordingly,  can be expanded as
\begin{align}
D_{CE}=\f{1}{M}\tr(\Iv-\Pv^T(\Pv\Pv^T+\bm{\Sigma_\eta})^{\dagger}\Pv).
\end{align}
Let $k_{CE}$ be the index $k$ that satisfies \eqref{eq:cedrf_central interval}: for $L\geq M$, we have for $k\leq k_{CE}$, $D_l=\theta$, and $1-2^{-2R_l}=0$; for $k>k_{CE}$, $D_l=\la_l+\sigma^2$, and $1-2^{-2R_l}=1-\theta/(\la+\sigma^2)$. Hence, for $k_{CE}\in[M]$, we have
\begin{align}
D_{CE}
&=\f{1}{M}{\rm Tr}(\Iv)-\f{1}{M}{\rm Tr}\lb\Pv\Pv^T\lb\Pv\Pv^T+\bm{\Sigma_\eta}\rb^{\dagger}\rb
\nonumber\\
&=1-\f{1}{M}{\rm Tr}\lb\Jv\Uv^T\Av\Av^T\Uv\Jv(\Jv\Uv^T\Av\Av^T\Uv\Jv+\sigma^2\Jv^2+\Jv\Dv)^{\dagger}\rb.
\end{align}
Recall that $\Uv$ is the orthogonal matrix that diagonalizes $\Av\Av^T$, we have $\Uv^T\Av\Av^T\Uv=\bm{\Lambda}\triangleq \diag([\la_1,\ldots,\la_r,0\ldots,0])$, and hence
\begin{align}
D_{CE}
&=1-\f{1}{M}{\rm Tr}\lb\Jv^2\bm{\Lambda}(\Jv^2\bm{\Lambda}+\sigma^2\Jv^2+\Jv\Dv)^{\dagger}\rb\nonumber\\
&=1-\f{1}{M}\sum_{l=1}^{M}\f{\la_l(1-2^{-2R_l})}{(\la_l+\sigma^2)(1-2^{-2R_l})+\min(\la_l+\sigma^2,\theta_k(\bm{\Sigma_Y},R))}.
\end{align}
Recall that, from the definition of $k_{CE}$, we have $\la_l+\sigma^2>\theta_k(\bm{\Sigma_Y},R)$ for $l\in[k_{CE}]$, and $\la_l+\sigma^2\leq \theta_k(\bm{\Sigma_Y},R)$ for $l>k_{CE}$.
Hence,
\begin{align}
D_{CE}
&=1-\f{1}{M}\lb\sum_{l=1}^{k_{CE}}\f{\la_l}{\la_l+\sigma^2}-\theta_k(\bm{\Sigma_Y},R) \sum_{l=1}^{k_{CE}}\f{\la_l}{(\la_l+\sigma^2)^2}\rb.
\end{align}
When $L<M$, for $k_{CE}\in[L]$, we have
\begin{align}
D_{CE}
&=1-\f{1}{M}\sum_{l=1}^{L}\f{\la_l(1-2^{-2R_l})}{(\la_l+\sigma^2)(1-2^{-2R_l})+D_l}\nonumber\\
&=1-\f{1}{M}\lb\sum_{l=1}^{k_{CE}}\f{\la_l}{\la_l+\sigma^2}-\theta_k(\bm{\Sigma_Y},R) \sum_{l=1}^{k_{CE}}\f{\la_l}{(\la_l+\sigma^2)^2}\rb,
\end{align}
which is the desired result.

\newpage
\section{Proof of Proposition \ref{prop:equality small rate}}
\label{app:performance gap}
%
The relationship between the functions $D_{CE}(R)$ and $D_{X|Y}(R)$  is better understood by explicitly showing that $D_{CE}(R)\geq D_{X|Y}(R)$. 
Fix $R$ and let $k_I$ and $k_{CE}$ be the indices that satisfy \eqref{eq:idrf linear interval} and \eqref{eq:cedrf_central interval}, respectively.
Next, if $k_I=k_{CE}=k$, we have
	\eas{
		D_{CE}(R)
		&=1-\f{1}{M}\sum_{l=1}^{k}\f{\la_l}{\la_l+\sigma^2}+\f{2^{-\f{2R}{k}}}{M}\lb\prod_{l=1}^{k}(\la_l+\sigma^2)\rb^{\f{1}{k}}\sum_{l=1}^{k}\f{\la_l}{(\la_l+\sigma^2)^2}\nonumber\\
		&=1-\f{1}{M}\sum_{l=1}^{k}\f{\la_l}{\la_l+\sigma^2}+\f{2^{-\f{2R}{k}}}{M}\lb\prod_{l=1}^{k}(\la_l+\sigma^2)\rb^{\f{1}{k}}k
		\lb \sum_{l=1}^{k} \f 1 k \f{\la_l}{(\la_l+\sigma^2)^2} \rb \nonumber\\
		%
		%
		& \geq 1-\f{1}{M}\sum_{l=1}^{k}\f{\la_l}{\la_l+\sigma^2}+\f{2^{-\f{2R}{k}}}{M}\lb\prod_{l=1}^{k}(\la_l+\sigma^2)\rb^{\f{1}{k}}k
		\lb \prod_{l=1}^{k} \f{\la_l}{(\la_l+\sigma^2)^2} \rb^{1/k}
		\label{eq:inequality means}
		\\
		& \geq 1-\f{1}{M}\sum_{l=1}^{k}\f{\la_l}{\la_l+\sigma^2}+\f{2^{-\f{2R}{k}}}{M} k
		\lb \prod_{l=1}^{k} \f{\la_l}{(\la_l+\sigma^2)} \rb^{1/k} \\
		&=D_{X|Y}(R),
	}{\label{eq:dce>=idrf}}
	where \eqref{eq:inequality means} follows from the AM-GM inequality so that equality is achieved only when $\la_l/(\la_l+\sigma^2)^2$ are equal for all $l\in[k]$.
	
	For the case $k_I\neq k_{CE}$, note that $k_I$ is the optimal number of active components for the optimal scheme, so that 
	\eas{
		D_{X|Y}(R)
		&\leq 1-\f{1}{M}\sum_{l=1}^{k_{CE}}
		\f{\la_l}{\la_l+\sigma^2}+\f{k_{CE}\cdot2^{-\f{2R}{k_{CE}}}}{M}\lb\prod_{l=1}^{k_{CE}}\frac{\la_l} {\la_l+\sigma^2}\rb^{\f{1}{k_{CE}}} \nonumber\\
		&\leq 1-\f{1}{M}\sum_{l=1}^{k_{CE}}\f{\la_l}{\la_l+\sigma^2}+\f{2^{-\f{2R}{k_{CE}}}}{M}\lb\prod_{l=1}^{k_{CE}}(\la_l+\sigma^2)\rb^{\f{1}{k_{CE}}}\sum_{l=1}^{k_{CE}}\f{\la_l}{(\la_l+\sigma^2)^2} \\
		&=D_{CE}(R).
	}{\label{eq:dce>=idrf 2}}
From \eqref{eq:dce>=idrf} and \eqref{eq:dce>=idrf 2} we realize that $D_{CE}(R)$ and $D_{X|Y}(R)$  are identical when  $\f{\la_l}{(\la_l+\sigma^2)^2}=\f{\la_1}{(\la_1+\sigma^2)^2}$
for all $l\in[k]$.
Note that the equation 
\ea{
\f{\la_k}{(\la_k+\sigma^2)^2}=\f{\la_1}{(\la_1+\sigma^2)^2}
}
has two solutions in $\la_k$ as in the RHS of \eqref{def:r0}.
%
%
Hence, for $r_0$ defined as in \eqref{def:r0}  we have that, if $R\leq \min\{R_{r_0+1}^{I},R_{r_0+1}^{CE}\}$, then $D_{X|Y}(R)=D_{CE}(R)$.
%
%

Finally, note that  if $r_0=L=M$, and $\f{\la_l}{(\la_l+\sigma^2)^2}=\f{\la_1}{(\la_1+\sigma^2)^2}$ for all $l$, then that $R_k^{I}=R_k^{CE}$ for all $k$ and 
 thus  $D_{X|Y}(R)=D_{CE}(R)$ for all values of $R$.
 %
%
%

\section{Proof of Theorem \ref{thm:performance gap}.}
\label{app:gap bound}
The proof relies on the following lemma from \cite{nguyen2008reversing}.
\begin{lem}\cite[Prop. 5]{nguyen2008reversing}
	Given a natural number $n$ greater than $1$, the smallest real number $k$ such that for all non-negative numbers $a_1,a_2,\ldots,a_n$ we have the inequality
	\ea{
		\f 1 n \sum_{i=1}^{n}a_i \leq  \lb \prod_{i=1}^n a_i \rb^{\f 1 n}  + k \max_{i \neq j} |a_i-a_j|,
	}
	is  $(n-1)/n$.
\end{lem}
Let $k_{CE}$ be the index $k$ that satisfies \eqref{eq:cedrf_central interval}.
Accordingly, the difference between the LHS and RHS of \eqref{eq:dce>=idrf 2} can be bounded as
\ea{
	& D_{CE}(R)-D_{X|Y}(R) \label{eq:gap} \\
	& =  \f{2^{-\f{2R}{k_{CE}}}}{M} \lb\prod_{l=1}^{k_{CE}}(\la_l+\sigma^2)\rb^{\f{1}{k_{CE}}}
	k_{CE}  \lb \f 1 { k_{CE} } \sum_{l=1}^{k_{CE}}\f{\la_l}{(\la_l+\sigma^2)^2} - \lb \prod_{l=1}^{k_{CE}}\f{\la_l} {(\la_l+\sigma^2)^2}\rb^{\f{1}{k_{CE}}}\rb \nonumber \\
	& \leq  \f{2^{-\f{2R}{k_{CE}}}}{M} k_{CE} \f {k_{CE}-1}{k_{CE}}  (\la_1+\sgs) \frac{1}{4\sigma^2} \nonumber \\
	& \leq  \f{2^{-\f{2R}{k_{CE}}}}{M}   k_{CE}  \f {\la_1+\sgs}{4\sgs} \nonumber \\
	&  \leq  \f{2^{-\f{2R}{L}}}{M}   L \f{\la_1+\sgs}{4\sgs}. \nonumber
}

\section{Proof of Theorem \ref{thm:performance gap 2}.}
\label{app:average loss}
%
%
%
Let us consider again the difference between optimal and CE performance in  \eqref{eq:gap} and write:
\begin{align}
&~~~D_{CE}(R)-D_{X|Y}(R)\nonumber\\
&\geq \frac{2^{-\frac{2R}{{k_{CE}}}}}{M}\left(\prod_{l=1}^{k_{CE}} (\la_l+\sigma^2)\right)^{\frac{1}{{k_{CE}}}}\cdot\left(\sum_{l=1}^{k_{CE}}\frac{\la_l}{(\la_l+\sigma^2)^2}-{k_{CE}}\left(\prod_{l=1}^{k_{CE}}\frac{\la_l}{(\la_l+\sigma^2)^2}\right)^{\frac{1}{{k_{CE}}}}\right)\nonumber\\	
&\geq
\frac{2^{-\frac{2R}{{k_{CE}}}}}{M}\left(\prod_{l=1}^{k_{CE}} (\la_l+\sigma^2)\right)^{\frac{1}{{k_{CE}}}}\cdot\left(\max_{l\leq {k_{CE}}}\sqrt{\frac{\la_l}{(\la_l+\sigma^2)^2}}-\min_{l\leq {k_{CE}}}\sqrt{\frac{\la_l}{(\la_l+\sigma^2)^2}}\right)^2\nonumber\\
&\geq\frac{\la_{{k_{CE}}+1}+\sigma^2}{M} \left(\max_{l\leq {k_{CE}}}\frac{\sqrt{\la_l}}{\la_l+\sigma^2}-\min_{l\leq {k_{CE}}}\frac{\sqrt{\la_l}}{\la_l+\sigma^2}\right)^2\nonumber\\
&\geq \frac{\la_{{k_{CE}}+1}+\sigma^2}{M}\left(\frac{(\la_{k_{CE}}+\sigma^2)^{k_{CE}}}{\prod_{l=1}^{{k_{CE}}}(\la_l+\sigma^2)}\right)^{\frac{1}{L}}\cdot\left(\max_{l\leq {k_{CE}}}\frac{\sqrt{\la_l}}{\la_l+\sigma^2}-\min_{l\leq {k_{CE}}}\frac{\sqrt{\la_l}}{\la_l+\sigma^2}\right)^2\nonumber\\
&\geq \frac{\la_{L}+\sigma^2}{M}2^{-\frac{2R}{L}}\left(\max_{l\leq {k_{CE}}}\frac{\sqrt{\la_l}}{\la_l+\sigma^2}-\min_{l\leq {k_{CE}}}\frac{\sqrt{\la_l}}{\la_l+\sigma^2}\right)^2.
\end{align}
The first inequality follows from the fact that $k_{CE}$ satisfies \eqref{eq:cedrf_central interval} but not necessarily \eqref{eq:idrf linear interval}, and hence could be
a non-optimal choice for the setting with full knowledge. The second inequality follows from a lower bound on the difference between AM and GM in \cite[Sec. II]{tung1975lower}. 
The third inequality follows from the definition of $k_{CE}$ as prescribed in \eqref{eq:cedrf_central interval}. 
%

For $R\leq R_2(\bm{\Sigma_{Y}})$, we have the obvious and trivial lower bound
\begin{align}
D_{CE}(R)-D_{X|Y}(R)\geq 0.
\end{align}
For $R>R_2(\bm{\Sigma_{Y}})$, we have
\begin{align}
\left(\max_{l\leq k}\frac{\sqrt{\la_l}}{\la_l+\sigma^2}-\min_{l\leq k}\frac{\sqrt{\la_l}}{\la_l+\sigma^2}\right)^2
\geq \left(\frac{\sqrt{\la_1}}{\la_1+\sigma^2}-\frac{\sqrt{\la_2}}{\la_2+\sigma^2}\right)^2,
\end{align}
hence
\begin{align}
D_{CE}(R)-D_{X|Y}(R)\geq  \frac{\la_{L}+\sigma^2}{M}\left(\frac{\sqrt{\la_1}}{\la_1+\sigma^2}-\frac{\sqrt{\la_2}}{\la_2+\sigma^2}\right)^2 2^{-\frac{2R}{L}}.
\end{align}

\end{document}